\documentclass[conference,letterpaper]{IEEEtran}

\usepackage{amsmath}
\usepackage{cases}
\usepackage{amsfonts}
\usepackage{amssymb}
\usepackage{boxedminipage}
\usepackage{graphicx}
\usepackage[usenames]{color}
\usepackage{array,color}
\usepackage{colortbl}
\usepackage{lineno}
\usepackage{slashbox,pict2e}
\usepackage{stfloats}
\usepackage{algorithm}
\usepackage{algorithmic}
\usepackage{multicol}
\usepackage{booktabs}
\usepackage{makecell}
\usepackage{bm}
\usepackage{float}

\begin{document}

\pagestyle{headings}
\pagenumbering{arabic}

\title{Tracking Angles of Departure and Arrival in a Mobile Millimeter Wave Channel}
\author{Chuang Zhang$^*$, ~Dongning Guo$^\dag$, ~Pingyi Fan$^*$ \\
$^*$Department of Electronic Engineering, Tsinghua University, Beijing, P.R. China\\
$^\dag$Department of Electrical Engineering and Computer Science, Northwestern University, Evanston, IL, USA\\
E-mail:~zhangchuang11@mails.tsinghua.edu.cn, ~dGuo@northwestern.edu, ~fpy@tsinghua.edu.cn}
\maketitle

\begin{abstract}
Millimeter wave provides a very promising approach for meeting the ever-growing traffic demand in next generation wireless networks. To utilize this band, it is crucial to obtain the channel state information in order to perform beamforming and combining to compensate for severe path loss. In contrast to lower frequencies, a typical millimeter wave channel consists of a few dominant paths. Thus it is generally sufficient to estimate the path gains, angles of departure (AoDs), and angles of arrival (AoAs) of those paths. Proposed in this paper is a dual timescale model to characterize abrupt channel changes (e.g., blockage) and slow variations of AoDs and AoAs. This work focuses on tracking the slow variations and detecting abrupt changes. A Kalman filter based tracking algorithm and an abrupt change detection method are proposed. The tracking algorithm is compared with the adaptive algorithm due to Alkhateeb, Ayach, Leus and Heath (2014) in the case with single radio frequency chain. Simulation results show that to achieve the same tracking performance, the proposed algorithm requires much lower signal-to-noise-ratio (SNR) and much fewer pilots than the other algorithm. Moreover, the change detection method can always detect abrupt changes with moderate number of pilots and SNR.
\end{abstract}


\section{Introduction} \label{sec_intro}

Millimeter wave communication is deemed as a promising technique for meeting the ever-increasing traffic demand in next generation wireless communication systems \cite{boccardi2014fdtd5g} \cite{andrews2014ww5gb}. Due to high attenuation loss inherent in this band, directional beamforming and combining should be applied to guarantee high enough signal-to-noise-ratio (SNR) for detection at receivers. Hence, it is necessary to obtain relatively accurate channel state information (CSI) at all transmitters and receivers.

Channel estimation in millimeter wave differs from that in lower frequencies in two major aspects: First, due to high cost and power consumption of A/D D/A converters with very high sampling rates, millimeter wave transmitters and receivers can only be equipped with limited number of radio frequency (RF) chains \cite{alkhateeb2014mpcsmms}. This imposes some constraints on the type of beamforming/combining that can be employed. Second, millimeter wave channel has limited scatterings due to large attenuation loss, high absorption loss, etc \cite{akdeniz2014mmwcmcce}. Therefore, it is sufficient to estimate channel parameters of these scattering paths instead of each element in the large channel matrix.

One straightforward approach to estimate the channel is to search the angles of departure (AoDs) and angles of arrival (AoAs) exhaustively, and use the direction with the largest gain as the beamforming/combining direction, as proposed in IEEE 802.11ad. However, the estimation accuracy of this approach is limited even with a large number of pilots. There are some other works which can improve the estimation accuracy, like \cite{alkhateeb2014cehpmwcs} \cite{lee2014essechmsmwc}. In \cite{alkhateeb2014cehpmwcs}, the authors proposed an adaptive algorithm to estimate the channel, which essentially searches the paths using beamforming vectors with different beamwidths in different stages. In \cite{lee2014essechmsmwc}, the authors formulated the channel estimation problem as a sparse signal recovery problem and solved it with the orthogonal matching pursuit (OMP) algorithm. Those algorithms either rely on transceivers with several RF chains or large number of quantization levels for the angles to achieve high estimation accuracy. For more practical transceivers with single RF chain and limited quantization levels of phase shifters, their improvements over the exhaustive search approach are limited.

\begin{figure}
  \centering
  \includegraphics[width=0.7\columnwidth]{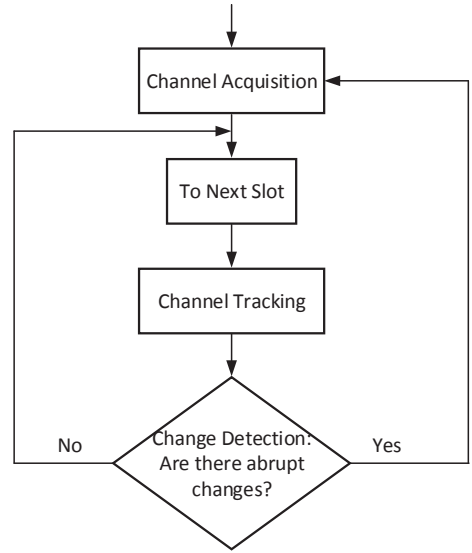}\\
  \caption{Channel acquisition, tracking and change detection system.}\label{fig_gen_procedure}
\end{figure}

This work studies the millimeter wave channel estimation problem by assuming a dual timescale channel variation model with both abrupt changes (e.g., blockage) and slow variations of AoDs and AoAs. The estimation framework is shown in Fig.~\ref{fig_gen_procedure}, where each time we detect an abrupt change, some channel acquisition method is used to obtain channel parameters, while if no abrupt change is detected, some channel tracking method is used to follow the slow variations. In this paper, we devise an efficient channel tracking algorithm based on Kalman filter and an abrupt change detection method. The channel acquisition problem is addressed in a separate paper. The proposed approach requires low SNR and low pilots overhead while maintaining a high tracking accuracy.

The rest of the paper is organized as follows. In Section \ref{sec_sysmod}, we present the system model. In Section \ref{sec_transpolicy}, the transmission scheme regarding sending pilots is discussed. In Section \ref{sec_kalmanfilter} and \ref{sec_acd}, we introduce the Kalman filter based tracking algorithm and the abrupt change detection method, respectively. In Section \ref{sec_simulation}, simulation results are presented. In Section \ref{sec_conclusion}, conclusions are drawn.

Throughout this paper, the following notations will be used. Matrices are denoted by bold uppercase letters (e.g., $\mathbf{A}$), vectors are denoted by bold lowercase letters (e.g., $\mathbf{a}$), scalars are denoted by lowercase letters (e.g., $a$). $\mathbf{A}^T$, $\mathbf{A}^*$, $\mathbf{A}^H$ denote the transpose, conjugate, Hermitian (conjugate transpose) of matrix $\mathbf{A}$, respectively. $\mathbf{I}_{N}$ denotes an $N\times N$ identity matrix. Finally, $[\mathbf{A};\mathbf{B}]$ denotes the matrix obtained by appending $\mathbf{B}$ under $\mathbf{A}$, where they must be of the same number of columns.

\section{System Model} \label{sec_sysmod}

\subsection{Structure of transmitter and receiver} \label{subsec_strtransreceiv}

\begin{figure}
  \centering
  \includegraphics[width=0.95\columnwidth]{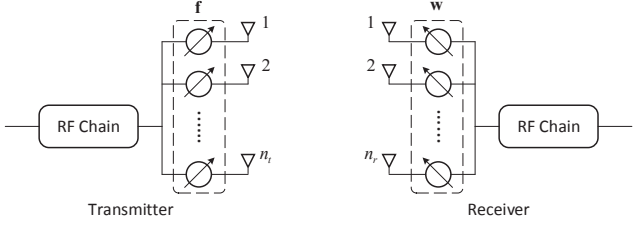}\\
  \caption{Transmitter and receiver.}\label{fig_txrx1RF}
\end{figure}

We assume that both the transmitter and the receiver utilize uniform linear arrays (ULAs), as shown by Fig. \ref{fig_txrx1RF}. The transmitter and the receiver have $n_t$, $n_r$ antennas, respectively. The separation between two antennas at the transmitter and receiver are $\Delta_t\lambda_c, \Delta_r\lambda_c$,  respectively, where $\lambda_c$ is the wavelength. Throughout the paper, we assume that the antennas are critically spaced, i.e., $\Delta_t=\Delta_r=\frac{1}{2}$. For ease of implementation, we consider that both the transmitter and the receiver have only one RF chain, hence, only analog beamforming/combining can be applied. Nevertheless, the proposed approach can be extended to the hybrid beamforming/combining scheme. We use $\mathbf{f}, \mathbf{w}$ to denote the beamforming vector, combining vector,  respectively, and
\begin{align}\label{eqn_bvector}
\mathbf{f}&=\tfrac{1}{\sqrt{n_t}}[e^{j\phi_1},e^{j\phi_2},\ldots,e^{j\phi_{n_t}}]^T, \\
\mathbf{w}&=\tfrac{1}{\sqrt{n_r}}[e^{j\psi_1},e^{j\psi_2},\ldots,e^{j\psi_{n_r}}]^T, \label{eqn_cvector}
\end{align}
where $\phi_i\in[0,2\pi], i=1,2,\ldots,n_t$, $\psi_i\in[0, 2\pi], i=1,2,\ldots,n_r$, i.e., each element in the beamforming and combining vectors are of constant modulus, and only the phase can be varied. In practice, the number of quantization levels of each phase shifter is usually limited, and we assume that each phase shifter can only have $N_t^{\text{max}}$, $N_r^{\text{max}}$ quantization levels at transmitters and receivers, respectively.

\subsection{Channel model} \label{subsec_chamod}

\begin{figure}
  \centering
  \includegraphics[width=0.95\columnwidth]{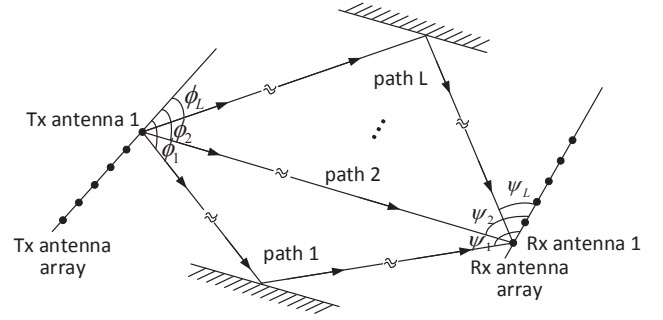}\\
  \caption{L-scatterer channel model.}\label{fig.chmod}
\end{figure}

We adopt the $L$-scatterer channel model, as in \cite{alkhateeb2014cehpmwcs} \cite{lee2014essechmsmwc}. Let $\phi_{l}, \psi_{l}$ be the AoD and AoA of path $l$, $l=1,2,\ldots,L$, and
\begin{flalign}  \label{eqn_AoDvec}
&\mathbf{e}_t(\phi)=\tfrac{1}{\sqrt{n_t}}[1,e^{-j\pi\cos{\phi}},\ldots,e^{-j\pi(n_t-1)\cos{\phi}}]^T,\\
&\mathbf{e}_r(\psi)=\tfrac{1}{\sqrt{n_r}}[1,e^{-j\pi\cos{\psi}},\ldots,e^{-j\pi(n_r-1)\cos{\psi}}]^T,\label{eqn_AoAvec}
\end{flalign}
then the $L$-scatterer channel can be expressed as \cite[P. 311]{tse2005fowc}
\begin{align}\label{eqn_chLscatterers}
\mathbf{H}=\sum_{l=1}^{L}\alpha_l\mathbf{e}_r(\psi_{l})\mathbf{e}_t^H(\phi_{l})
\end{align}
where $\alpha_l=\rho_l\sqrt{n_tn_r}e^{-j\frac{2\pi d_{l}}{\lambda_c}}$, and $\alpha_l$, $\rho_l$, $d_l$ are the path gain, attenuation, distance between transmit antenna $1$ and receive antenna $1$ along path $l$, respectively.

Let $\bm{\alpha}=[\alpha_1,\ldots,\alpha_L]^T$ and $\bm{\theta}=[\phi_{1},\ldots,\phi_{L},\psi_{1},\ldots,\psi_{L}]^T$. Then the path gain vector $\bm{\alpha}$ and the angle vector $\bm{\theta}$ fully determine the channel.

\subsection{Channel variation model} \label{subsec_chavarmod}

In practice, paths in millimeter wave channel are mostly influenced by two factors: 1) abrupt changes (disappearance or appearance of dominant paths) due to sudden environmental change, like blockage by vehicles or pedestrians; 2) noise disturbance to AoDs and AoAs, like slow rotations of hand devices or vibrations of base station poles \cite{hur2013mmbwbascn}. To capture the impact of these two factors, we adopt a simple channel model with variations on two timescales. First, we denote the time period between two consecutive abrupt changes as a block, and its length is variable, for instance, we can assume it follows geometric distribution. In different blocks, either some existing dominant paths vanish or new dominant paths appear. Second, within a block, the AoDs and AoAs of existing dominant paths vary slowly due to noise and other impairments, while gains of these paths remain invariant. Such variations happen in a much shorter timescale and we denote it as a slot. To keep tracking the channel, it is necessary to estimate the channel on these two timescales. On the block scale, we should determine when abrupt changes occur and then in the beginning of each block, acquire the dominant paths. On the slot scale, with the initial estimate of the channel, we should keep tracking these paths in each slot. Since these abrupt changes take place infrequently, we focus this work on the tracking problem and the abrupt change detection problem assuming an initial estimate of the CSI. The channel acquisition problem is addressed separately.

The path gain vector $\bm{\alpha}$ is assumed to remain constant in different slots of the same block, while the angle vector $\bm{\theta}$ varies following
\begin{align} \label{eqn_anglevary}
\bm{\theta}(n)=\mathbf{A}\bm{\theta}(n-1)+\mathbf{u}(n),
\end{align}
where $n$ denotes the $n$th slot, $\mathbf{A}$ is a known $2L\times 2L$ matrix, $\mathbf{u}(n)$ is Gaussion noise and $\mathbf{u}(n)\sim \mathcal{N}(0,\mathbf{Q}_u)$.

\section{Transmission Scheme} \label{sec_transpolicy}

We discuss how to send pilots in this section. Suppose the transmitter sends symbol $x$ using beamforming vector $\mathbf{f}$, and the receiver combines with vector $\mathbf{w}$. Then the observation at the receiver is
\begin{align} \label{eqn_observation}
y=&\mathbf{w}^H\mathbf{H}\mathbf{f}x+\mathbf{w}^H\mathbf{v},
\end{align}
where $\mathbf{v}$ is Gaussian noise and $\mathbf{v}\sim \mathcal{CN}(0,\sigma_v^2\mathbf{I}_{n_r})$.

\subsection{Beamforming and combining vectors}

In general, we can use any beamforming and combining vectors of forms (\ref{eqn_bvector}) and (\ref{eqn_cvector}), respectively. However, considering the structure of the channel (\ref{eqn_chLscatterers}), we further restrict our beamforming and combining vectors to be of forms (\ref{eqn_AoDvec}) and (\ref{eqn_AoAvec}), respectively. Then the directions of corresponding vectors are determined by one parameter.

Suppose we use $\mathbf{f}=\mathbf{e}_t(\bar{\phi})$, $\mathbf{w}=\mathbf{e}_r(\bar{\psi})$ as the beamforming and combining vector, respectively. Then it can be calculated that the beamforming and combining gain of path $l$ are
\begin{flalign}\label{eqn_fsample}
&\mathbf{e}_t^H(\phi_{l})\mathbf{e}_t(\bar{\phi})=\frac{1}{n_t}\frac{1-e^{j\pi n_t(\cos{\phi_l}-\cos{\bar{\phi}})}}{1-e^{j\pi (\cos{\phi_l}-\cos{\bar{\phi}})}},\\
&\mathbf{e}_r^H(\bar{\psi})\mathbf{e}_r(\psi_{l})
=\frac{1}{n_r}\frac{1-e^{-j\pi n_r(\cos{\psi_l}-\cos{\bar{\psi}})}}{1-e^{-j\pi (\cos{\psi_l}-\cos{\bar{\psi}})}}, \label{eqn_wsample}
\end{flalign}
respectively.

The AoD $\phi_{l}$ is in the range $[0,2\pi]$, however, since $\cos{\phi_l}$ is symmetric around $\pi$, we only need to consider the range $[0, \pi]$. For instance, if $\phi_l=\frac{4\pi}{3}$ and the estimate is  $\frac{\pi}{3}$, then based on (\ref{eqn_fsample}), the estimate is accurate because the gain is the same whether the beamforming direction is $\frac{4\pi}{3}$ or $\frac{\pi}{3}$. Similarly, it is only necessary to consider the range $[0,\pi]$ for $\psi_l$.

To detect the AoDs, we select $N_{t}\leq N_t^{\text{max}}$ beamforming vectors to cover the range $[0,\pi]$. Essentially, we are quantizing the AoD range $[0, \pi]$ into $N_t$ bins, and there can be many ways to conduct quantization. Here we quantize the range of $\cos\phi_l$, i.e., $[-1,1]$ uniformly and then use the angle corresponding to the center of each bin as the beamforming direction.

The transmitter sends pilots using beamforming vectors $\mathbf{f}_p=\mathbf{e}_t(\bar{\phi}_{p}), p=1,2,\ldots,N_{t}$, in turn, and for each beamforming vector, the receiver uses combining vectors $\mathbf{w}_q=\mathbf{e}_r(\bar{\psi}_{q}), q=1,2,\ldots,N_{r}$, in turn. Consequently, we search all $N_{r}\times N_{t}$ directions and obtain $N_{r}\times N_{t}$ received symbols.

\subsection{Observation}

When the $p$th beamforming vector is sent and the $q$th combing vector is used, the received symbol is
\begin{align} \label{eqn_obValue}
y_{qp}=\mathbf{w}_q^H\mathbf{H}\mathbf{f}_p+\mathbf{w}_q^H\mathbf{v},
\end{align}
where we set symbol $x=1$ compared with Eqn. (\ref{eqn_observation}).

Then the observation matrix is
\begin{align}
\mathbf{Y}=\left[\begin{array}{cccc}
y_{11}&y_{12}&\cdots&y_{1N_{t}}\\
y_{21}&y_{22}&\cdots&y_{2N_{t}}\\
\vdots&\vdots&\ddots&\vdots\\
y_{N_{r}1}&y_{N_{r}2}&\cdots&y_{N_{r}N_{t}}
\end{array}
\right].
\end{align}

Let $\mathbf{W}=[\mathbf{w}_1,\mathbf{w}_2,\ldots,\mathbf{w}_{N_{r}}]$, $\mathbf{F}=[\mathbf{f}_1,\mathbf{f}_2,\ldots,\mathbf{f}_{N_{t}}]$, then the observed data can be written as
\begin{align} \label{eqn_obserMatrix}
\mathbf{Y}=\mathbf{W}^H\mathbf{H}\mathbf{F}+\mathbf{V},
\end{align}
where $\mathbf{V}$ is an $N_{r}\times N_{t}$ Gaussian noise matrix and each element is independent identically distributed (i.i.d.).

Let $g_{qp}(\bm{\theta})=\mathbf{w}_q^H\mathbf{H}\mathbf{f}_p$, and by setting $\Omega_t^{lp}=\cos{\phi_{l}}-\cos{\bar{\phi}_{p}}$, $\Omega_r^{lq}=\cos{\psi_{l}}-\cos{\bar{\psi}_{q}}$, it can be expressed as
\begin{align} \label{eqn_gtheta}
g_{qp}(\bm{\theta})= \sum_{l=1}^L\frac{\alpha_l}{n_t n_r}\frac{1-e^{-j\pi n_r\Omega_r^{lq}}}{1-e^{-j\pi \Omega_r^{lq}}}\frac{1-e^{j\pi n_t\Omega_t^{lp}}}{1-e^{j\pi \Omega_t^{lp}}}.
\end{align}

Then the observation matrix becomes
\begin{align}
\mathbf{Y}=\left[\begin{array}{cccc}
g_{11}(\bm{\theta})&g_{12}(\bm{\theta})&\cdots&g_{1N_{t}}(\bm{\theta})\\
g_{21}(\bm{\theta})&g_{22}(\bm{\theta})&\cdots&g_{2N_{t}}(\bm{\theta})\\
\vdots&\vdots&\ddots&\vdots\\
g_{N_{r}1}(\bm{\theta})&g_{N_{r}2}(\bm{\theta})&\cdots&g_{N_{r}N_{t}}(\bm{\theta})
\end{array}
\right]+\mathbf{V}.
\end{align}

We vectorize the observation matrix by concatenating the columns of $\mathbf{Y}$. Let $\mathbf{y}=\text{vec}(\mathbf{Y})$, $\mathbf{v}=\text{vec}(\mathbf{V})$ and
\begin{align} \label{eqn_gfunc}
\mathbf{g}(\bm{\theta})=&[g_{11}(\bm{\theta}),\ldots,g_{N_{r}1}(\bm{\theta}),g_{12}(\bm{\theta})\ldots,g_{N_r2}(\bm{\theta}),\ldots,\nonumber\\
&g_{1N_{t}}(\bm{\theta}),\ldots,g_{N_{r}N_{t}}(\bm{\theta})]^T.
\end{align}
Then
\begin{align} \label{eqn_vecobserve}
\mathbf{y}=\mathbf{g}(\bm{\theta})+\mathbf{v},
\end{align}
where $\mathbf{v}\sim \mathcal{C}N(0,\mathbf{Q}_v)$, and $\mathbf{Q}_v=\sigma^2_v\mathbf{I}_{N_{r}N_{t}}$.
Obviously, $\mathbf{g}(\bm{\theta})$ is not linear with respect to (w.r.t.) $\bm{\theta}$.

\section{Kalman filter based channel tracking} \label{sec_kalmanfilter}

\begin{figure*}[!hb]
\hrulefill
\begin{align} \label{eqn_partial_tl}
\frac{\partial{g}_{qp}(\bm{\theta})}{\partial{\phi_{l}}}&=\frac{-\alpha_l\sin{\phi_{l}}}{n_t n_r}\frac{1-e^{-j\pi n_r\Omega_r^{lq}}}{1-e^{-j\pi \Omega_r^{lq}}}\frac{j\pi  e^{j\pi \Omega_t^{lp}}-j\pi n_t e^{j\pi n_t \Omega_t^{lp}}+j\pi(n_t-1)e^{j\pi(n_t+1)\Omega_t^{lp}}}{(1-e^{j\pi\Omega_t^{lp}})^2}, \\
\frac{\partial{g}_{qp}(\bm{\theta})}{\partial{\psi_{l}}}&=\frac{-\alpha_l\sin{\psi_{l}}}{n_t n_r}\frac{-j\pi  e^{-j\pi  \Omega_r^{lq}}+j\pi n_r e^{-j\pi n_r \Omega_r^{lq}}-j\pi(n_r-1)e^{-j\pi(n_r+1)\Omega_r^{lq}}}{(1-e^{-j\pi\Omega_r^{lq}})^2} \frac{1-e^{j\pi n_t\Omega_t^{lp}}}{1-e^{j\pi \Omega_t^{lp}}}, \label{eqn_partial_rl}
\end{align}
\end{figure*}

Based solely on observation (\ref{eqn_vecobserve}), classical estimation algorithms like least squares can be used. However, since $\mathbf{g}(\bm{\theta})$ is nonlinear w.r.t. $\bm{\theta}$  (here $\bm{\alpha}$ is assumed known by some channel acquisition algorithm), these algorithms usually require grid search with very high computational complexity. On the other hand, if we consider both the channel variation model (\ref{eqn_anglevary}) and the observation (\ref{eqn_vecobserve}), we find that they fit into Kalman filter (\cite[Chap. 13]{kay2009estimation}) framework. Since Kalman filter provides a recursive way to update the estimate and is computationally more efficient, we employ this method for the tracking problem.

Kalman filter requires that both the signal evolution and the observation be linear with the signal. However, in our case, the observation $\mathbf{y}$ is nonlinear w.r.t. the angle vector $\bm{\theta}$ in (\ref{eqn_vecobserve}). Thus, we should first use a linear approximation for the observation. Define $\bm{\hat{\theta}}(n|n-1)$ as the linear minimum mean square error (MMSE) estimator of $\bm{\theta}(n)$ based on $\{\mathbf{y}(0), \mathbf{y}(1),\ldots, \mathbf{y}(n-1)\}$, then
\begin{align} \label{eqn_linearappro}
\mathbf{y}(n)=&\mathbf{g}\big(\bm{\theta}(n)\big)+\mathbf{v}(n) \nonumber\\
=&\mathbf{g}\big(\bm{\theta}(n)-\bm{\hat{\theta}}(n|n-1)+\bm{\hat{\theta}}(n|n-1)\big)+\mathbf{v}(n)\nonumber\\
\approx &\mathbf{g}\big(\bm{\hat{\theta}}(n|n-1)\big)+\nonumber\\
& \frac{\partial \mathbf{g}(\bm{\theta}(n))}{\partial \bm{\theta}(n)}|_{\bm{\theta}(n)=\bm{\hat{\theta}}(n|n-1)}\Big(\bm{\theta}(n)-\bm{\hat{\theta}}(n|n-1)\Big)+\mathbf{v}(n)\nonumber\\
=&\mathbf{C}(n)\bm{\theta}(n)+\mathbf{v}(n)+\mathbf{d}(n)
\end{align}
where $\mathbf{C}(n)=\frac{ \partial\mathbf{g}(\bm{\theta}(n))}{\partial \bm{\theta}(n)}|_{\bm{\theta}(n)=\bm{\hat{\theta}}(n|n-1)}$ is the partial derivative vector of $\mathbf{g}(\bm{\theta}(n))$ w.r.t. $\bm{\theta}(n)$ and its calculation is shown in (\ref{eqn_partial_tl}) and (\ref{eqn_partial_rl}),  $\mathbf{d}(n)=\mathbf{g}\big(\bm{\hat{\theta}}(n|n-1)\big)-\mathbf{C}(n)\bm{\hat{\theta}}(n|n-1)$.

Then with the channel evolution model (\ref{eqn_anglevary}) and the observation model (\ref{eqn_linearappro}), we can use Kalman filter for channel tracking. In particular, the algorithm is shown as Algorithm~\ref{alg_kfct}. The notations are: $\widehat{\bm{\theta}}(n|i)$ is the linear MMSE estimator of $\bm{\theta}(n)$ based on $\{\mathbf{y}(0), \mathbf{y}(1),\ldots, \mathbf{y}(i)\}$, $\mathbf{M}(n|n-1)=\mathbb{E}\big((\bm{\theta}(n)-\widehat{\bm{\theta}}(n|n-1))(\bm{\theta}(n)-\widehat{\bm{\theta}}(n|n-1))^H\big)$  is the minimum prediction mean square error (MSE) matrix, $\mathbf{M}(n|n)=\mathbb{E}\big((\bm{\theta}(n)-\widehat{\bm{\theta}}(n|n))(\bm{\theta}(n)-\widehat{\bm{\theta}}(n|n))^H\big)$ is the MMSE matrix, $\mathbf{K}(n)$ is the Kalman gain matrix, $\widehat{\bm{\theta}}(0|0)$ and $\mathbf{M}(0|0)$ are the initial values of $\widehat{\bm{\theta}}(n|n)$ and $\mathbf{M}(n|n)$, respectively.
\begin{algorithm}
\caption{Kalman Filter based Channel Tracking}
\label{alg_kfct}
\begin{algorithmic}[1]
\STATE \textbf{Initialization:} $\widehat{\bm{\theta}}(0|0)\leftarrow\bm{\theta}(0), \mathbf{M}(0|0)\leftarrow \mathbf{0}$,
\WHILE{slot $n$ still in the same block}
\STATE Prediction:
 $\widehat{\bm{\theta}}(n|n-1) \leftarrow \mathbf{A}\widehat{\bm{\theta}}(n-1|n-1)$,
\STATE Minimum Prediction MSE:
 $\mathbf{M}(n|n-1) \leftarrow \mathbf{A}\mathbf{M}(n-1|n-1)\mathbf{A}^H+\mathbf{Q}_u$,
\STATE Kalman Gain Matrix:
$\mathbf{K}(n) \leftarrow \mathbf{M}(n|n-1)\mathbf{C}^H(n)\Big(\mathbf{Q}_v+\mathbf{C}(n)\mathbf{M}(n|n-1)\mathbf{C}^H(n)\Big)^{-1}$,
\STATE Correction:
$\widehat{\bm{\theta}}(n|n)\leftarrow \widehat{\bm{\theta}}(n|n-1)+\mathbf{K}(n)\Big(\mathbf{y}(n)-\mathbf{g}\big(\widehat{\bm{\theta}}(n|n-1)\big)\Big)$,
\STATE \textbf{Output $\widehat{\bm{\theta}}(n|n)$ as the estimate of $\bm{\theta}(n)$},
\STATE MMSE:
$\mathbf{M}(n|n)\leftarrow \Big(\mathbf{I}-\mathbf{K}(n)\mathbf{C}(n)\Big)\mathbf{M}(n|n-1)$,
\STATE $n\leftarrow n+1$,
\ENDWHILE
\end{algorithmic}
\end{algorithm}

Algorithm \ref{alg_kfct} provides a recursive way to estimate the channel, so each time only the current observation $\mathbf{y}(n)$ and the previous estimate $\widehat{\bm{\theta}}(n-1|n-1)$ need to be stored.

One thing concerning the implementation of Algorithm~\ref{alg_kfct} needs to be mentioned here. Note that $\bm{\theta}(n)\in \mathcal{R}^{2L}$ is always real while $\mathbf{y}(n)$, $\mathbf{C}(n)$, $\mathbf{v}(n)$ and $\mathbf{d}(n)$ in (\ref{eqn_linearappro}) are all complex. Then each time we update $\widehat{\bm{\theta}}(n|n)$ in Step 6 of Algorithm \ref{alg_kfct} would result in complex values for $\widehat{\bm{\theta}}(n|n)$.   To deal with this, we let  $\widetilde{\mathbf{y}}(n)=[\text{Re}(\mathbf{y}(n)); \text{Im}(\mathbf{y}(n))]$, $\widetilde{\mathbf{C}}(n)=[\text{Re}(\mathbf{C}(n));\text{Im}(\mathbf{C}(n))]$,  $\widetilde{\mathbf{v}}(n)=[\text{Re}(\mathbf{v}(n));\text{Im}(\mathbf{v}(n))]$, $\widetilde{\mathbf{d}}(n)=[\text{Re}(\mathbf{d}(n));\text{Im}(\mathbf{d}(n))]$, and formulate an equivalent expression of (\ref{eqn_linearappro}) as
\begin{align}\label{eqn_linearapproReal}
\widetilde{\mathbf{y}}(n)=\widetilde{\mathbf{C}}(n)\bm{\theta}(n)+\widetilde{\mathbf{v}}(n)+\widetilde{\mathbf{d}}(n).
\end{align}

By applying Algorithm \ref{alg_kfct} to (\ref{eqn_anglevary}) and (\ref{eqn_linearapproReal}), we are dealing directly with real vectors and $\widehat{\bm{\theta}}(n|n)$ is always real.

\section{Abrupt Change Detection} \label{sec_acd}

In this section, we develop a method to detect abrupt channel changes based on the Kalman filter tracking algorithm.

The two hypotheses at any time $n$ are:

$\mathcal{H}_0$: there is no abrupt change in the channel;

$\mathcal{H}_1$: there are abrupt changes in the channel.

We use $\mathbf{y}(n)$ for such a test and define
\begin{flalign} \label{eqn_abtchdet}
L(\mathbf{y}(n))= (\mathbf{y}(n)-\mathbf{g}(\hat{\bm{\theta}}(n)))^H \mathbf{Q}_v^{-1}(\mathbf{y}(n)-\mathbf{g}(\hat{\bm{\theta}}(n))),
\end{flalign}
where $\hat{\bm{\theta}}(n)$ is the estimate of $\bm{\theta}$ using Kalman filter (Algorithm \ref{alg_kfct}). Conceivably, if the channel acquisition and tracking are accurate enough, $L(\mathbf{y}(n))$ would be small in the case without abrupt changes since the main term remained in $\mathbf{y}(n)-\mathbf{g}(\hat{\bm{\theta}}(n))$ is noise according to (\ref{eqn_vecobserve}). On the other hand, if there are abrupt changes, the remained term would include gains of some dominant paths, which would be large.

We then decide $\mathcal{H}_1$ if
\begin{align} \label{eqn_Lgamma}
L(\mathbf{y}(n))>\gamma,
\end{align}
where $\gamma$ is a predefined threshold. To decide the threshold $\gamma$, the false alarm probability can be used, which is defined as
\begin{align}\label{eqn_fa}
P_{\text{FA}}=P\{L(\mathbf{y}(n))>\gamma;\mathcal{H}_0\}.
\end{align}

We can determine an approximate expression for $\gamma$ given a $P_{\text{FA}}$. With relatively accurate channel acquisition and tracking, $2L(\mathbf{y}(n))$ would follow Chi-Squared distribution with degree $2N_tN_r$, i.e., $2L(\mathbf{y}(n))\sim \chi_{2N_tN_r}^2$, where $\chi_{\nu}^2$ denotes Chi-Squared distribution with degree $\nu$. Let $Q_{\chi_{\nu}^2}(x)$ be the right-tail probability for a $\chi_{\nu}^2$ random variable, i.e., $Q_{\chi_{\nu}^2}(x)=\int_x^\infty p(t)\mathbf{d}t$, then given a false alarm probability, $P_{\text{FA}}$, we can determine a corresponding threshold
\begin{align} \label{eqn_thres}
\gamma=\frac{1}{2}Q_{\chi_{2N_rN_t}^2}^{-1}(P_{\text{FA}}).
\end{align}
We will show that this approximate threshold works well for the proposed framework in the simulations.

The detection method is presented as Algorithm \ref{alg_GLRTacd}.
\begin{algorithm}
\caption{Abrupt Change Detection}
\label{alg_GLRTacd}
\begin{algorithmic}[1]
\STATE Given a $P_{\text{FA}}$, determine the threshold $\gamma$ using (\ref{eqn_thres}),
\WHILE{True}
\STATE In slot $n$, use Algorithm \ref{alg_kfct} to obtain $\hat{\bm{\theta}}(n)$,
\STATE Determine whether there are abrupt changes based on (\ref{eqn_abtchdet}) and (\ref{eqn_Lgamma}),
\ENDWHILE
\end{algorithmic}
\end{algorithm}

\section{Simulation Results} \label{sec_simulation}

\begin{figure}
  \centering
  \includegraphics[width=0.92\columnwidth]{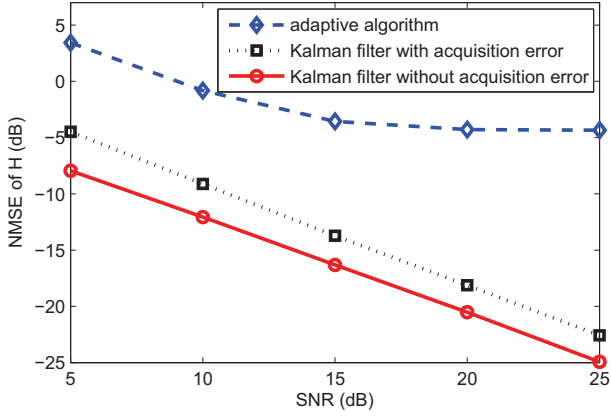}\\
  \caption{NMSE of $\mathbf{H}$ versus SNR, $N_t=N_r=16$, $\sigma_u^2=(\frac{0.5}{180}\pi)^2$. }\label{fig_NMSEvsSNR}
\end{figure}

\begin{figure}
  \centering
  \includegraphics[width=0.92\columnwidth]{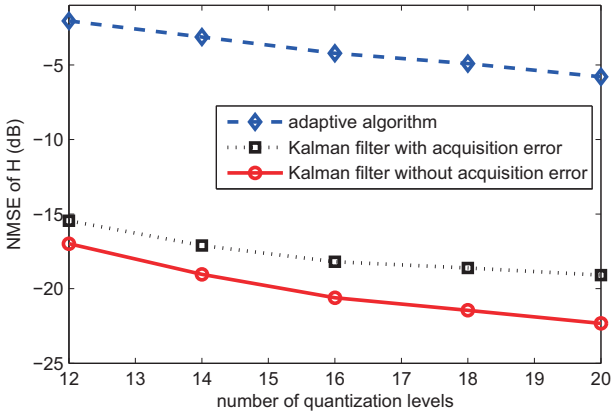}\\
  \caption{NMSE of $\mathbf{H}$ versus number of quantization levels, SNR=$20$ dB, $\sigma_u^2=(\frac{0.5}{180}\pi)^2$.}\label{fig_NMSEvspilots}
\end{figure}

\begin{figure}
  \centering
  \includegraphics[width=0.92\columnwidth]{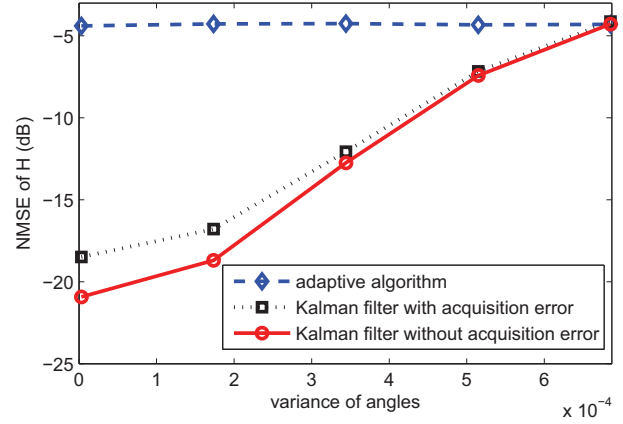}\\
  \caption{ NMSE of $\mathbf{H}$ versus channel variations, SNR=$20$ dB, $N_t=N_r=16$. }\label{fig_NMSEvsvariation}
\end{figure}

\begin{figure*}[!htb]
 \centering
  \includegraphics[width=2\columnwidth, height=5cm]{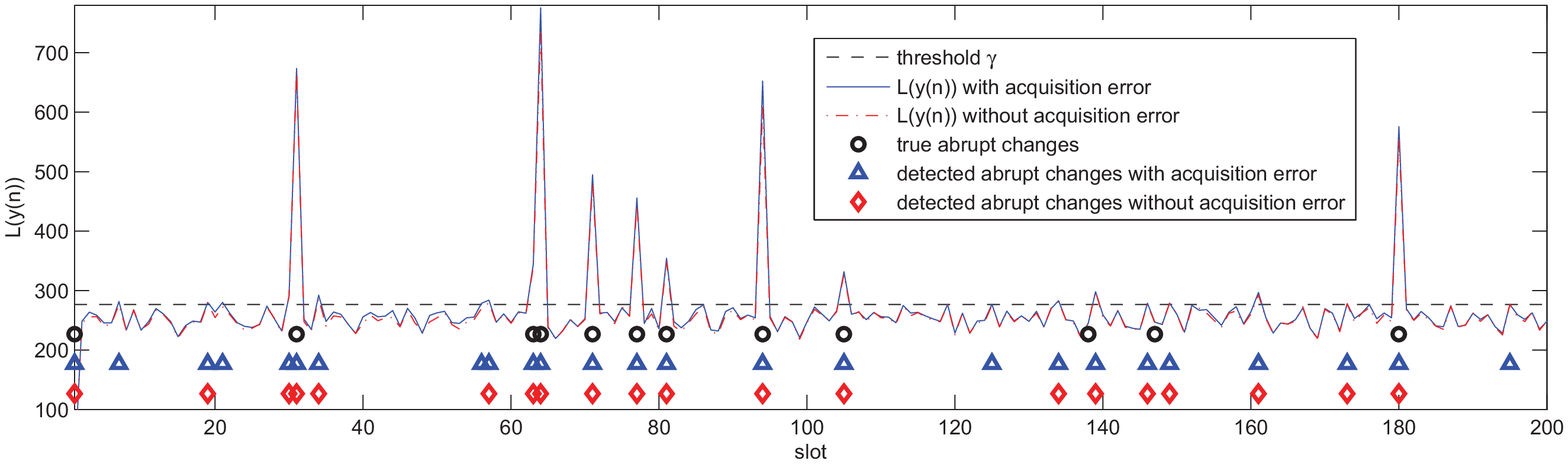}\\
  \caption{Change detections, SNR=$20$ dB, $N_t=N_r=16$, $\sigma_u^2=(\frac{0.5}{180}\pi)^2$, $p_{\text{app}}=0.0254$, $p_{\text{dis}}=0.0127$, $p_{\text{FA}}=0.1$. }\label{fig_change_detection}
\end{figure*}

\begin{table}
\centering
\caption{Common Simulation parameters}
\label{tab_compars}
\begin{tabular}{c|c|c|c|c|c|c}
  \Xhline{1.2pt}
  \Gape[1.5ex]{$n_t$} & $n_r$  & $L$ & $\mathbf{A}$ & $\mathbf{Q}_u$ & $\mathbf{Q}_v$  & SNR \\
  \hline
   \Gape[1.5ex]{16} & 16  & $3$ & $\mathbf{I}_{2L}$ & $\sigma_u^2\mathbf{I}_{2L}$ & $\sigma_v^2\mathbf{I}_{N_{r}N_{t}}$ & $10\log_{10}\frac{n_tn_r}{\sigma_v^2}$ \\
  \Xhline{1.2pt}
\end{tabular}
\end{table}

We show the performance of the proposed approach in this section. Common simulation parameters are given in Table \ref{tab_compars}. For ease of simulation, we set $\mathbf{Q}_u=\sigma_u^2\mathbf{I}_{2L}$ in the table, which means that the variations of AoDs and AoAs of different paths are independent. It can be shown that the proposed algorithm applies for the general correlated case as well. Besides, $\sigma_u^2$ determines how fast the channel varies, so we can change this parameter to simulate different channel variation speeds.

We compare the Kalman filter algorithm (Algorithm \ref{alg_kfct}) with the adaptive algorithm in \cite{alkhateeb2014cehpmwcs}. In the case with single RF chain at both transmitters and receivers, the adaptive algorithm is similar to the OMP algorithm in \cite{lee2014essechmsmwc}. As the performance measure, we use the normalized mean square error (NMSE) of $\mathbf{H}$, which is defined as $\frac{\mathbb{E}(||\widehat{\mathbf{H}}-\mathbf{H}||^2_F)}{\mathbb{E}(||\mathbf{H}||^2_F)}$, where $\mathbf{H}$ is the true channel matrix, $\widehat{\mathbf{H}}$ is the estimated channel matrix and $||\star||_F$ is the Frobenius norm.

Figs. \ref{fig_NMSEvsSNR} - \ref{fig_NMSEvsvariation} show the tracking performances of these algorithms versus SNR, number of quantization levels and channel variations, respectively. For each point in the figures, we simulate $1000$ blocks, with each block lasting $100$ slots. In the first slot of each block, we generate $L$ paths with gains following $\mathcal{CN}(0, n_tn_r)$, and angles following $\mathcal{U}(0,\pi)$. Then in the next $99$ slots, the path gains do not change while the angles evolve according to (\ref{eqn_anglevary}). For the Kalman filter algorithm, we simulate two cases, with one having acquisition error and the other not, where the acquisition error $\bm{\alpha}-\hat{\bm{\alpha}}$ follows $\mathcal{CN}(0,\sigma_v^2\mathbf{I}_L)$. The adaptive algorithm estimates $\bm{\alpha}$ and $\bm{\theta}$ in each slot. The channel variation speed parameter $\sigma_u^2$ is unknown for all algorithms. In the Kalman filter algorithm, we use $(\frac{2}{180}\pi)^2$ as the guessed value instead of its real value. The adaptive algorithm does not need this parameter.

As shown by Figs. \ref{fig_NMSEvsSNR} and \ref{fig_NMSEvspilots}, both the Kalman filter algorithm and the adaptive algorithm would achieve higher tracking accuracy with the increase of SNR or number of quantization levels. Besides, the Kalman filter algorithm has superior performance over the adaptive algorithm, and the increase of estimation accuracy is always larger than $10$ dB without acquisition error. Even with acquisition error, the Kalman filter still performs much better than the adaptive algorithm.

Fig. \ref{fig_NMSEvsvariation} illustrates that the performance of the Kalman filter algorithm would deteriorate when the channel variation speed increases, which mainly arises from applying Kalman filter for nonlinear observation. With large variation noise, the Kalman filter algorithm loses track, and estimation error can accumulate gradually. So there is a threshold of channel variation speed for this algorithm. However, determining the threshold can be difficult analytically and is out of the scope of this paper. In the simulations, when $\sigma_u^2\leq(\frac{1}{180}\pi)^2$, the Kalman filter algorithm can track the channel accurately. When $\sigma_u^2=(\frac{1}{180}\pi)^2$, the average drift of the angle ($\mathbb{E}(|u|)$) would be $\sqrt{\frac{2}{\pi}}\sigma_u\approx 0.798$. Suppose each slot lasts $1$ ms, then the angle variation speed is $798^\circ/s$, which is equivalent to rotating a mobile device more than two rounds per second. Thus, $\sigma_u^2=(\frac{1}{180}\pi)^2$ is actually a very fast angle variation speed and for any reasonable speed, the algorithm can be applied.

Fig. \ref{fig_change_detection} shows the change detection results of Algorithm~\ref{alg_GLRTacd}. The channels are generated in the following way. In the first slot, we generate $L$ paths independently, with their path gains following $\mathcal{CN}(0, n_tn_r)$ and angles following $\mathcal{U}(0, \pi)$. Then in the following slots, for each path, we independently decide whether it disappears or appears by tossing a coin. If in the previous slot, the path exists, then the disappearance probability of the path in the current slot is $p_{\text{dis}}$. If in the previous slot, the path does not exist, then the appearance probability of the path in the current slot is $p_{\text{app}}$. If the path is not subject to an abrupt change, then the path gain does not change while the angles evolve according to (\ref{eqn_anglevary}). As can be seen from this figure, Algorithm \ref{alg_GLRTacd} detect all true abrupt changes whether acquisition error exists or not. On the other hand, the false alarm probability is higher with acquisition error than that without acquisition error, which is $\frac{13}{200}$ versus $\frac{7}{200}$ in these $200$ slots. So in order to use the tracking algorithm and the change detection method, a relatively accurate channel acquisition algorithm is desired.

\section{Conclusion} \label{sec_conclusion}

In this paper, we have investigated the millimeter wave channel estimation problem. A dual timescale channel variation model is assumed to characterize abrupt changes and slow variations. Then a Kalman filter based tracking algorithm and an abrupt change detection method have been proposed. Simulation results showed that both the tracking algorithm and the abrupt change detection method work quite well in the proposed framework if the AoDs and AoAs vary with a reasonable speed. In addition, the performance of the proposed schemes degrades gracefully with acquisition error, which allowed for extensions by jointly considering channel acquisition, tracking and abrupt change detection.

\section*{Acknowledgement}

This work was partly supported by the China Major State Basic Research Development Program (973 Program) No.~2012CB316100(2), National Natural Science Foundation of China No.~61201203. The work of D. Guo was supported in part by a gift from Futurewei Technologies and by the National Science Foundation under Grant No.~ECCS-1231828.

\bibliographystyle{IEEEtran}
\bibliography{mmwave}

\begin{thebibliography}{1}
\providecommand{\url}[1]{#1}
\csname url@samestyle\endcsname
\providecommand{\newblock}{\relax}
\providecommand{\bibinfo}[2]{#2}
\providecommand{\BIBentrySTDinterwordspacing}{\spaceskip=0pt\relax}
\providecommand{\BIBentryALTinterwordstretchfactor}{4}
\providecommand{\BIBentryALTinterwordspacing}{\spaceskip=\fontdimen2\font plus
\BIBentryALTinterwordstretchfactor\fontdimen3\font minus
  \fontdimen4\font\relax}
\providecommand{\BIBforeignlanguage}[2]{{%
\expandafter\ifx\csname l@#1\endcsname\relax
\typeout{** WARNING: IEEEtran.bst: No hyphenation pattern has been}%
\typeout{** loaded for the language `#1'. Using the pattern for}%
\typeout{** the default language instead.}%
\else
\language=\csname l@#1\endcsname
\fi
#2}}
\providecommand{\BIBdecl}{\relax}
\BIBdecl

\bibitem{boccardi2014fdtd5g}
F.~Boccardi, R.~W. {Heath Jr.}, A.~Lozano, T.~L. Marzetta, and P.~Popovski,
  ``Five disruptive technology directions for {5G},'' \emph{IEEE Commun. Mag.},
  vol.~52, no.~2, pp. 74--80, Feb. 2014.

\bibitem{andrews2014ww5gb}
J.~G. Andrews, S.~Buzzi, W.~Choi, S.~V. Hanly, A.~Lozano, A.~C.~K. Soong, and
  J.~C. Zhang, ``What will {5G} be?'' \emph{IEEE J. Sel. Areas Commun.},
  vol.~32, no.~6, pp. 1065--1082, Jun. 2014.

\bibitem{alkhateeb2014mpcsmms}
A.~Alkhateeb, J.~Mo, N.~G. Prelcic, and R.~W. {Heath Jr.}, ``{MIMO} precoding
  and combining solutions for millimeter-wave systems,'' \emph{IEEE Commun.
  Mag.}, vol.~52, no.~12, pp. 122--131, Dec. 2014.

\bibitem{akdeniz2014mmwcmcce}
M.~R. Akdeniz, Y.~Liu, M.~K. Samimi, S.~Sun, S.~Rangan, T.~S. Rappaport, and
  E.~Erkip, ``Millimeter wave channel modeling and cellular capacity
  evaluation,'' \emph{IEEE J. Sel. Areas Commun.}, vol.~32, no.~6, pp.
  1164--1179, Jun. 2014.

\bibitem{alkhateeb2014cehpmwcs}
A.~Alkhateeb, O.~E. Ayach, G.~Leus, and R.~W. {Heath Jr.}, ``Channel estimation
  and hybrid precoding for millimeter wave cellular systems,'' \emph{IEEE J.
  Sel. Topics Signal Process.}, vol.~8, no.~5, pp. 831--846, Oct. 2014.

\bibitem{lee2014essechmsmwc}
J.~Lee, G.~Gil, and Y.~H. Lee, ``Exploiting spatial sparsity for estimating
  channels of hybrid {MIMO} systems in millimeter wave communications,'' in
  \emph{Proc. IEEE Global Commun. Conf.}, Dec. 2014, pp. 3326--3331.

\bibitem{tse2005fowc}
D.~N.~C. Tse and P.~Viswanath, \emph{Fundamentals of Wireless
  Communication}.\hskip 1em plus 0.5em minus 0.4em\relax Cambridge University
  Press, 2005.

\bibitem{hur2013mmbwbascn}
S.~Hur, T.~Kim, D.~J. Love, J.~V. Krogmeier, T.~A. Thomas, and A.~Ghosh,
  ``Millimeter wave beamforming for wireless backhaul and access in small cell
  networks,'' \emph{IEEE Trans. Commun.}, vol.~61, no.~10, pp. 4391--4403, Oct.
  2013.

\bibitem{kay2009estimation}
S.~M. Kay, \emph{Fundamentals of Statistical Signal Processing, Volume I:
  Estimation Theory}.\hskip 1em plus 0.5em minus 0.4em\relax Prentice Hall PTR,
  Jan. 2009.

\end{thebibliography}

\end{document}